# Dynamic particle packing in freezing colloidal suspensions

Jiaxue You[1], Jincheng Wang[1], Lilin Wang[2], Ziren Wang[3], Zhijun Wang[1*], Junjie Li[1] and Xin Lin[1*]

1-State Key Laboratory of Solidification Processing, Northwestern Polytechnical University, Xi'an 710072, P. R. China

2-School of Materials Science and Engineering, Xi'an University of Technology, Xi'an 710048, P. R. China

3-Department of Physics, Chongqing University, Chongqing 401331, P. R. China

**Abstract**: In the field of freezing colloidal suspensions, it is important to understand the particle-scale behavior of particle packing. Here, we reveal the dynamics of particle packing by identifying the behavior of each single particle in situ. The typical pattern consists of locally ordered clusters and amorphous defects. The microscopic mechanism of pattern formation is ascribed to the non-equilibrium particle-packing process on the particle scale, described with the Péclet number. The macroscopic migration of a particle layer is also revealed by an analytical model involving parameters of freezing speed and initial volume fraction of particles.

Freezing of colloidal suspensions is ubiquitous in nature and technology [1]. It is an important factor in many research areas, such as ice templating of porous ceramics [2], polymers and composites [3], bone tissue engineering [4], science of soft matter [5], geophysical science [6], thermal energy storage [7], crystal growth [8], cryobiology [9], etc. In all of these cases, the segregation of particles from the growing ice and the consequent increase of particle concentration in the fluid regions are vital. In particular, the arrangement of segregated particles on the scale of single particles is important, because the particle-scale structure is the key to understanding the rejection of particles and hence predicting the large-scale structure [10].

During freezing of a suspension, particles are expelled from ice [11,12], forming a close-packed layer in front of the freezing interface. This process can cause self-assembly of the particles [13] similar to that caused by drying [14,15] or sedimentation [16] of colloidal suspensions. Similarities between these patterns suggest that the physics underlying the colloidal behavior may be similar, though the driving forces in each case differ. Therefore, knowledge gained from studying particle packing in freezing colloidal suspensions may be applicable to colloidal

---

*Corresponding author. Tel.:86-29-88460650; fax: 86-29-88491484
 E-mail address: zhjwang@nwpu.edu.cn (Zhijun Wang), xlin@nwpu.edu.cn (Xin Lin)

suspensions in diverse circumstances.

Researchers have tried to reveal dynamic particle packing during freezing colloidal suspensions through theoretical models [17,18], experiments [2,10,13,19-27] and simulations [28], but are still far from a complete understanding of the phenomenon [26,27]. Presently, there is no theory that can fully predict the morphology or detailed characteristics of particle packing. Previous theories assume that the particles in the condensed layer form a random close packing [17,18], and the condensed layer is considered uniform with an average particle density. However, the detailed structure and dynamics of the condensed layer has not previously been identified by experiments on the scale of individual particles. Most studies involve a posteriori analysis of samples after fixing the particle structure [2,22-25]. They provided only static information about the final arrangement of particles. Some experiments have tried to resolve the dynamic behavior of particle layers by using X-ray radiography and tomography [19-21,26,27] as well as small-angle X-ray scattering [10]. X-ray techniques can probe inside visibly opaque materials. X-ray tomography can even provide a full three-dimensional reconstruction of the samples. However, none of these techniques provided information about the dynamic packing status on the scale of individual particles due to limited spatial resolution (>1 μm) in fast X-ray tomography (temporal resolution <1 s) [29]. Molecular dynamics simulations [28] have been also used to investigate dynamic particle packing during freezing suspensions, which stimulates us to conduct laboratory experiments to gain information about how real systems behave.

In this Letter, we present in-situ observation of dynamic particle packing on the scale of individual particles made during directional freezing in carefully controlled experiments. The typical pattern in the close-packed particle layer consists of locally ordered clusters and amorphous defects. The microscopic mechanism of pattern formation is investigated by particle packing process on the particle scale. Finally, the mechanisms of macroscopic particle layer migration are quantified by an analytical model involving parameters of freezing speed and initial volume fraction of particles.

In our experiments, a narrow-gapped sample cell was designed to obtain quasi-two-dimensional, mono-layer suspensions of micron-sized spherical particles in which the packing behavior of individual particles could be identified. The manufacturing and operation of the colloidal-monolayer sample cells is identical to Ref.[30]. The glass surfaces were rigorously cleaned so that particles did not stick to the walls. The colloidal suspensions were brushed onto a glass slide and another glass slide was used to seal the suspensions. Finally, edges of the slides were glued together with Araldite to form a sample cell of fixed thickness. The gap ($\approx$2.6 microns) between the slides was a little larger than the diameter ($\approx$1.73 microns) of the particles to allow Brownian motion of the monolayer of particles. The particles floated around midplane of the cell because the density of particles (1.03 g/cm$^3$) matches with water (1 g/cm$^3$) very well. A high-precision, directional-solidification apparatus was used to freeze the colloidal suspensions, and a CCD camera with 2580 × 1944 sensitive elements on a time-lapse video recorder was used to make continuous recordings [31]. The large optical contrasts between the areas of ice, close-packed particles and suspensions allowed a simple image-analysis technique to be employed. We used suspensions of polystyrene microspheres (PS) approximating hard-sphere interactions, which is an ideal system to investigate the freezing of colloidal suspensions [32]. The mean diameter of PS particles used was d=1.73 μm, with poly-dispersity smaller than 5% (Bangs Lab, USA). The thickness of the sample cell was around 2.6 microns, which is confirmed by Supplementary Movie S1. The general criterion of the thickness is the out-plane fluctuation along the direction of thickness [30]. The length and width of the sample cell were approximately 80,000 microns and 20,000 microns respectively, which are about four orders of magnitude larger than the thickness. The particle packing density $n$, defined as particle number per μm$^2$ (i.e., two-dimensional area number density), in the quasi two-dimensional thin-film suspensions, is used to represent the volume fraction of particles.

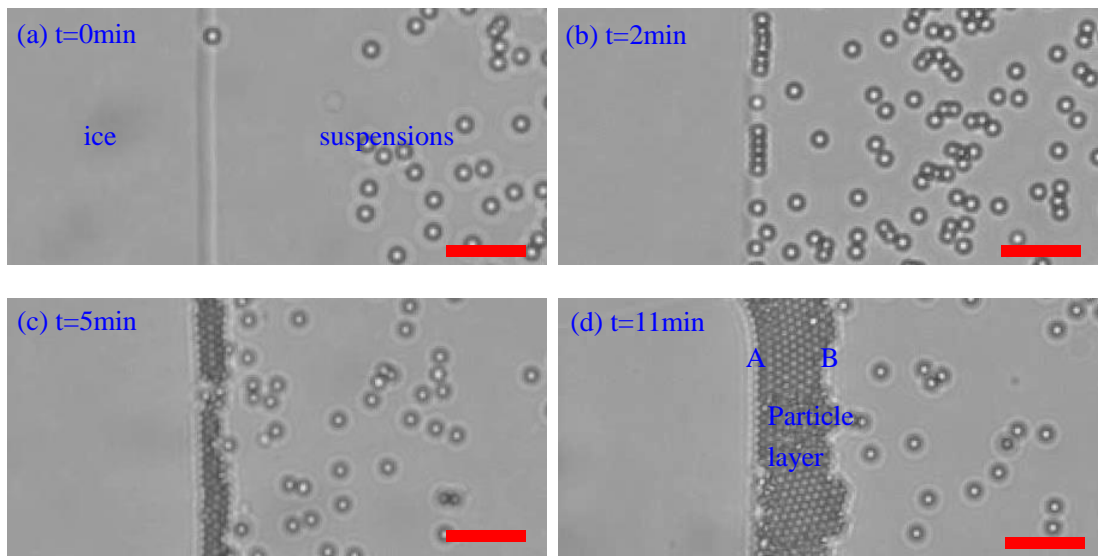

Fig.1 Dynamic creation of a close-packed particle layer during directional freezing of colloidal suspensions with a thermal gradient G=9.24 K/cm and pulling speed V=0.4 μm/s from right to left. The left side of the sample is the cooling zone, while the right side of the sample is the heating zone. A is freezing interface of ice and B is packing interface of particles. The scale bars are 10μm.

Figure 1 shows the dynamic process of creating a close-packed particle layer during directional freezing. The continuous dynamic process is shown in the Supplementary Movie S2. In Fig.1, the left side of the sample is the cooling zone, while the right side is the heating zone, which builds a linear thermal gradient G=9.24 K/cm. Directional freezing is provided by sample translation from right to left [31]. When the pulling speed, V=0.4 μm/s, is applied, the freezing interface starts advancing to catch up with the pulling speed. Once the freezing interface encounters particles at these low growth speeds, the particles are pushed ahead by the interface [11], leading to an increasing particle packing density ahead of the freezing interface. The particles firstly form a line at the interface, as shown in Fig.1(b). Particles are continually attached into the first line of particles with the movement of freezing interface (Fig.1(c)). Eventually, a close-packed particle layer forms, as shown in Fig.1(d). The thickness of the particle layer increases with the attachment of particles, and the packing interface of particles moves forward.

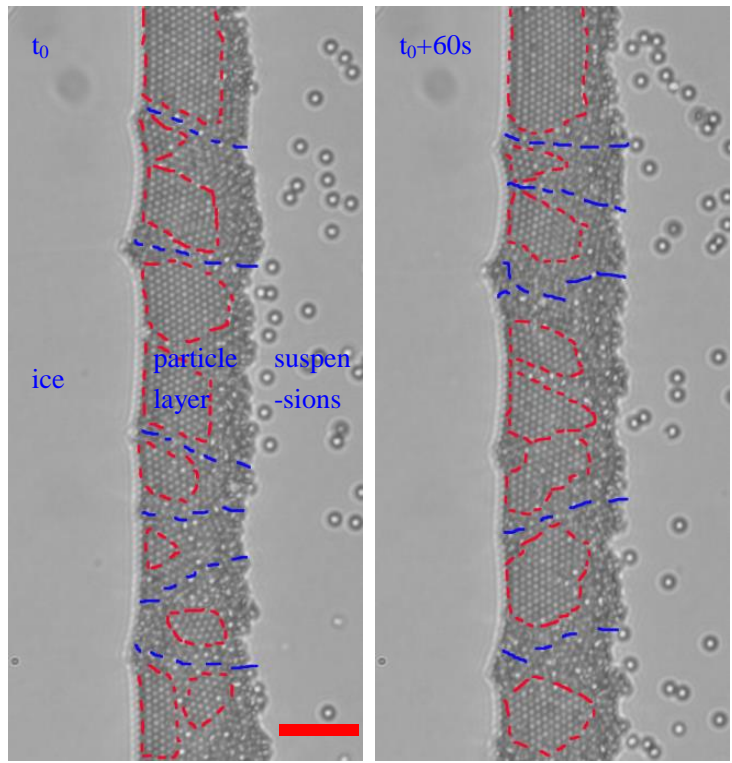

Fig.2 Dynamic evolution of the particle layer in freezing PS suspensions at different moments. The locally ordered clusters are circled by red dash lines. The defects are marked by blue dash lines. V=0.4μm/s, G=9.24K/cm. The scale bar is 10 μm.

Figure 2 shows the details of the close-packed particle layer. During the accumulation of particles towards the close-packed particle layer, the packing interface of particles is always rough, similar to the non-faceted solidification of metals [33]. Behind the rough packing interface of particles, locally ordered clusters appear, and some amorphous defects also exist between these clusters. This packing is similar to nano-crystalline (i.e., the locally ordered clusters,) and their grain boundaries (i.e., the defects). In Fig.2, the locally ordered clusters are marked as red dashed lines, while the amorphous defects are marked as blue dashed lines.

The formation of locally ordered clusters in Fig.2 is a volume-dependent colloidal crystallization. With the growth of freezing ice, the particles are expelled into water so that the volume fraction of particles in front of the freezing interface increases, leading to entropic ordering of particles [30,34]. Therefore locally ordered clusters appear at the freezing interface. Along with the drying of colloidal suspensions [14,15] and sedimentation of colloidal suspensions [16], the freezing of colloidal suspensions may be used as another method of colloidal crystallization which may arouse a general interest of colloidal science and material design.

The formation of amorphous defects is usually related to many factors, but the major factor in this system is non-equilibrium particle packing caused by growth speed of the close-packed particle layer. There are competitions between Brownian motion and the movement of the packing interface of the particle layer [35]. Brownian motion tends to make the particle packing at the interface move towards equilibrium, while the rapid movement of the packing interface may provide less relaxation time and drives the system out of equilibrium packing. If the growth speed of the particle layer is small enough, the packing system is near equilibrium, which is beneficial to entropic ordering of particles. On the contrary, the high growth speed of the particle layer leads to non-equilibrium packing of particles. This implies that a faster growth speed of the particle layer will induce more amorphous defects.

Other minor factors, such as poly-dispersity of particle size etc., are also considered in the formation of amorphous defects. When packing of particles with various sizes, the poly-dispersity of particle size can cause amorphous packing [36,37]. However, the particles we used had a poly-dispersity smaller than 5%, while amorphous packing requires a poly-dispersity of particle size larger than the range 6 to 12% [36,37]. Accordingly, the poly-dispersity of present particle size (<5%) has a limited effect on the formation of amorphous defects [37]. Therefore, non-equilibrium particle packing should be the major factor in the formation of the amorphous defects we observed.

From the qualitative analysis above, a semi-quantitative characterization can also be given as follows. The formation of amorphous defects is due to the non-equilibrium particle packing. The relaxation from Brownian motion can further promote the annihilation of defects. This relaxation can be characterized by the Brownian time, $\tau_B = d^2/D$, where $D = \mu k_B T$ is diffusion coefficient of particles in a confined space. $k_B$ is Boltzmann's constant, T the absolute temperature, μ the mobility of particles in a confined space and d the diameter of particles. The non-equilibrium of particle packing can be characterized by the time that the packing interface moves forward the distance of one particle diameter, $\tau_P = d/V_p$, where $V_p$ is the growth speed of particle layer. The competition between the non-equilibrium particle packing and relaxation can be normalized as Péclet number [14,16].

$$P_e = \frac{\tau_B}{\tau_P} = \frac{dV_p}{D}, \quad (1)$$

where Pe is Péclet number. A bigger Pe indicates more non-equilibrium of the packing system.

To confirm the semi-quantitative characterization of the non-equilibrium particle packing, i.e., the speed-dependent amorphous defects, three different pulling speeds, V=0.2 μm/s, 0.4 μm/s and 1.6 μm/s, under the same initial particle packing density, $n_0$=1.46% μm$^{-2}$, are applied respectively, as shown in Fig.3. It is easy to understand that the growth speed of the particle layer is provided by the movement of growing ice which is proportional to the pulling speed with regard to a constant $n_0$. To quantitatively describe the amount of defects, particle packing density in the particle layer $n_p$ is used, since a smaller $n_p$ implies more amorphous defects in the hard sphere packing. The $n_p$ have been averagely counted as 38.5±0.4%μm$^{-2}$, 36±0.3%μm$^{-2}$ and 34±0.4%μm$^{-2}$, under V=0.2 μm/s, 0.4 μm/s and 1.6 μm/s, respectively. For the present experiments, $\tau_B$ is almost constant in the confined particle layer, while $\tau_P \approx$ 8.6s, 4.3s and 1.1s for V=0.2 μm/s, 0.4 μm/s and 1.6 μm/s, respectively. Hence, Pe increases with the increase of pulling speed, corresponding to the decreasing $n_p$. The relevance of the variations indicates that the amorphous defects are determined by the non-equilibrium particle dynamic packing. The simulation work of Barr et al. [28] also showed the balance between diffusion velocity of particles and interface displacement velocity of ice determines particle packing status which is consistent with the present analysis and experiments.

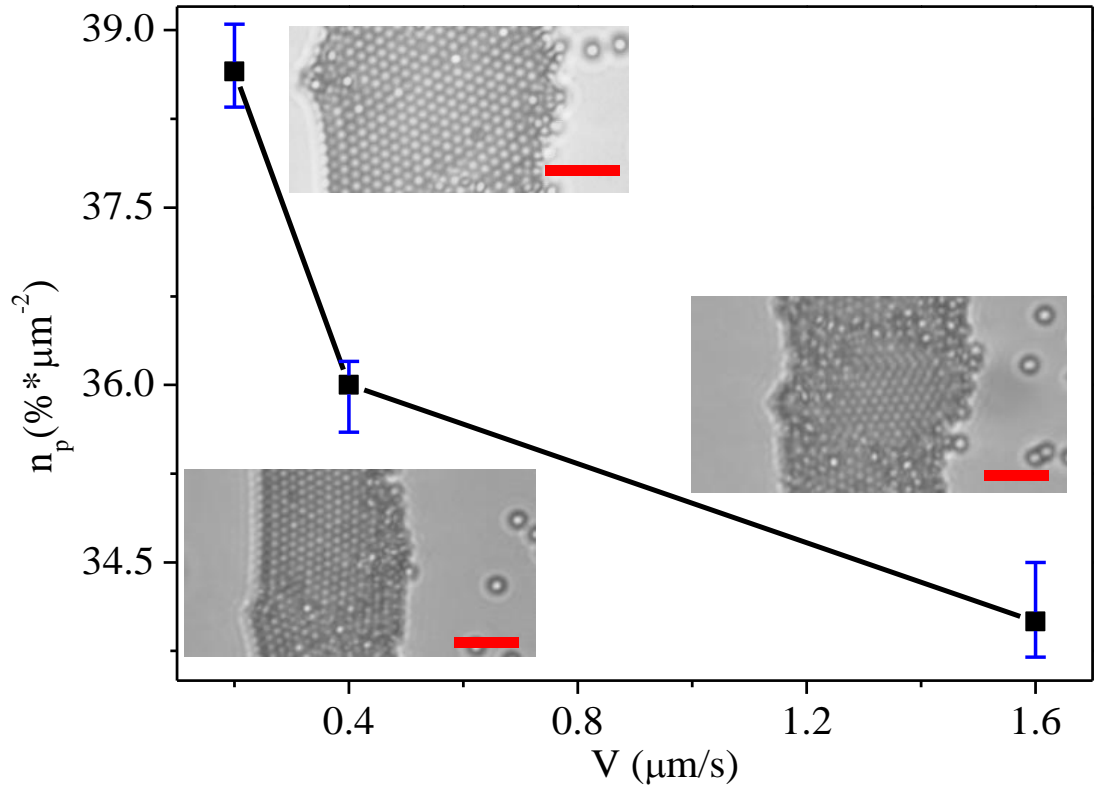

Fig.3 the particle layer under different pulling speeds with the constant thermal gradient G=9.24K/cm and particle density $n_0$=1.46%μm$^{-2}$. $n_p$= 38.5±0.4%μm$^{-2}$, 36±0.3%μm$^{-2}$ and 34±0.4%μm$^{-2}$, under pulling speeds V=0.2 μm/s, 0.4 μm/s and 1.6 μm/s, respectively. The insets are experimental particle packing under different pulling speeds. The scale bar is 6 μm.

We have confirmed the pattern of particle packing and its formation mechanism. These findings shed light on physical mechanisms of pattern formation in freezing of colloidal suspensions. On the particle scale, the pattern of particle packing in the close-packed layer includes ordered and disordered clusters. It is obvious that the local particle packing density depends on the packing status. With the proof of the amorphous defects here, it is possible that the growth of pore ice gives priority to the area of defects in ice banding formation [38].

The pattern in the particle layer is also dynamically evolving, compared the pattern at different moments, as shown in Fig.2. The dynamic evolution is like grain-boundary sliding and grain rotation [39]. The dynamic evolution of the particle layer is considered as the synthesis results of the propelling force of freezing interface on the particle layer and viscous flow of water in the particle layer. Other minor drag forces, such as friction between glass surfaces and particles in the present experiments, can be added into the viscous force of water flow. The assessment of the friction is given in the Supplementary Text S3. The growing ice pushes the whole particle layer forward. The propelling force causes stress in the particle layer. The stress transmission, through the whole particle layer, brings the dynamic evolution of the locally ordered clusters and the amorphous defects. On the other hand, the particles are also dragged by water during exchanging their positions with water ahead of the freezing ice.

The above discussions expose the formation mechanism of the close-packed particle layer on micro-scale. There is also another important issue on macro-scale, i.e., interactions between the freezing ice and the close-packed particle layer [19]. The experimental results also provide quantitative information about the interactions.

In the macro scale, there are two interfaces appearing in the freezing colloidal suspensions, i.e., the freezing interface and the packing interface of particles. Figure 4 shows the evolution speeds of these two interfaces and their relationship.

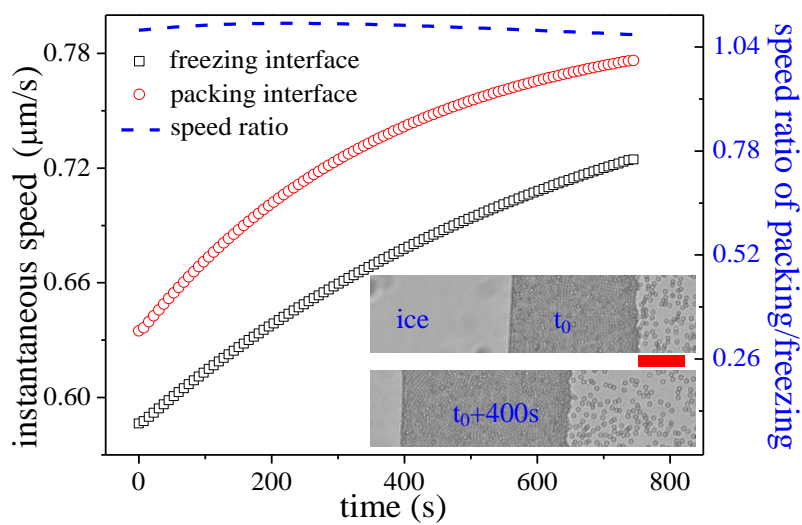

Fig.4 instantaneous interface speeds of freezing ice and packing particles. V=0.8μm/s, G=9.24K/cm, $n_0$=2.92%μm$^{-2}$. The insets are the freezing interface and the particle layer at different moments. The scale bar is 20μm. The directional freezing is provided by sample translation from right to left, resulting in freezing interface growing from left to right in the laboratory frame. However, speed of freezing interface slightly lags behind pulling speed, which makes a displacement of the freezing interface to the cold side under fixed field of view.

Inset of Fig. 4 shows the planar freezing interface pushes away the particle layer smoothly, so the pulling speed V=0.8 μm/s is smaller than both $V_c$ (critical speed of particle engulfment) and $V_s$ (critical speed of solutal diffusion instability). The width of the particle layer H increases with time and can be calculated as $H=(V_p-V_i)t$. The $V_p$ and $V_i$ represent respectively interface speeds of close-packed particle layer and freezing ice. Figure 4 shows the dynamic evolutions of $V_p$ and $V_i$, under $n_0=2.92\%\mu m^{-2}$. Compared with the growth of the freezing interface, the growth of the particle layer is a litter faster, resulting in the increasing width of the particle layer. Both the interface speeds of freezing ice and particle layer are gradually increasing to keep up with the pulling speed. Interestingly, the speed ratio of particle layer and freezing ice keeps constant. The mean value of speed ratio is 1.090±0.008. To explore the physical significance of this constant speed ratio, the derivation of the relation between interface speeds of particle layer and freezing ice is proposed as:

$$V_p = \frac{n_p}{n_p-n_0}V_i = \frac{1}{1-n_0/n_p}V_i. \qquad (2)$$

Note that the volume change in the water/ice phase transformation is neglected. Equation (2) shows that the speed of the particle layer is always faster than the speed of growing ice, unless $n_0 = 0$. The speed ratio $\frac{1}{1-n_0/n_p}$ can be used to calculate the particle packing density at the freezing interface, as $n_p=35.45\%\mu m^{-2}$, for the present condition. Furthermore, Eq.(2) can be utilized to examine the present statistics of particle packing density.

According to Fig.4 and Eq.(2), there are close connections between the freezing ice and the close-packed particle layer. On the one hand, the $V_i$ partially determines the $V_p$. The increasing $V_i$ leads to a bigger $V_p$ which will cause a smaller $n_p$, due to the non-equilibrium packing of particles as mentioned above. The smaller $n_p$ will also result in a bigger $V_p$, according to Eq.(2). Thus, the increasing $V_i$ may induce a nonlinear increase of $V_p$. On the other hand, the close-packed particles preclude the speed of freezing ice to keep up with the pulling speed, due to the added resistance of the accumulated particles on the flow of water feeding ice growth. Therefore, the interface speed of freezing ice always lags behind the pulling speed in the stage of initial transient, as shown in Fig.4. After this stage, the growth of the particle layer may lead to an increasing resistance due to the accumulated particles. Hence, the flow of water into the freezing ice may be more difficult, resulting in the continuous decreasing speed of ice [38]. As to the effect of $n_0$, according to Eq.(2), $V_p$ is proportional to $n_0$, which implies that a bigger $n_0$ will result in a larger $V_p$ with the same $V_i$, leading to more amorphous defects in the particle layer.

**Conclusions**

In the present work, the dynamic establishing of a close-packed particle layer in freezing of colloidal suspensions is in-situ observed. The pattern of the close-packed particle layer is identified as locally ordered clusters and amorphous inter-defects. The formation mechanism of amorphous defects is mainly from the competition between the particles' Brownian motion and the dynamic particle attachment on the particle layer interface, and the dimensionless Péclet number is used to describe the dynamic packing behavior. The macroscopic migration of packing interface of particles is determined by the freezing process and the initial particle volume fraction, which is quantitatively revealed by an analytical model. In the future, this quasi two-dimensional thin film suspensions system with directional freezing can facilitate more quantitative investigations on the ice-templating method [2], thermal regelation of particle clusters [40] and deformation of non-crystalline materials [41].

**Acknowledgements**

J.Y. thanks Grae Worster for his revising the manuscript and numerous discussions. This research has been supported by Nature Science Foundation of China (Grant Nos. 51371151 and 51571165), Free Research Fund of State Key Laboratory of Solidification Processing (100-QP-2014 and 158-QP-2016), the Fund of State Key Laboratory of Solidification Processing in NWPU (13-BZ-2014) and the Fundamental Research Funds for the Central Universities (3102015ZY020).